\documentclass[12pt]{iopart}

\usepackage{graphicx}
\usepackage{cite}
\begin{document}

\title{pQCD Calculations of Heavy Quark and $J/\psi$ Production}

\author{Matteo Cacciari}

\address{LPTHE\\
Universit\'e P. et M. Curie - Paris 6, France}
\ead{cacciari@lpthe.jussieu.fr}
\begin{abstract}
We review the present status of theoretical predictions for 
both closed ($J/\psi$) and open heavy quark production in high energy collisions,
and their comparisons to experimental data.

\end{abstract}


\section{Introduction}

The aim of this talk is to review the theoretical predictions of
perturbative QCD (pQCD) for the production of either closed ($J/\psi$) or
open heavy quark states in high energy collisions, and to compare them to the
available experimental measurements. While the focus of Quark
Matter is of course on heavy ion collisions, we shall however only describe 
predictions for the far simpler case of hadron collisions, aiming at
establishing benchmarks for more complex environments. A further restriction 
is to consider only collinear factorization, since it is in this framework 
that calculations are most  developed and, by
often being carried out beyond leading order, they allow for an estimate of
their own theoretical uncertainty.

\section{$J/\psi$ production}

The literature on heavy quarkonium production is extremely vast and it is 
impossible
to fully cover it in a short talk. Hence the decision to describe only a  specific
approach, and to select a few processes able to highlight both its strengths
and its weaknesses (for more comprehensive reviews see 
\cite{Kramer:2001hh,Brambilla:2004wf,Lansberg:2006dh}).

In heavy quarkonium physics one immediately recognizes three energy scales, the
heavy quark mass $m$, its momentum $mv$ and its kinetic energy $mv^2$, which
allow to separate the problem into different parts. In particular, the
relatively large scale of the heavy quark mass allows one to identify a short
distance, perturbative component, controlled by the value of the strong
coupling: $\alpha_s(m_c) \simeq 0.35$ and $\alpha_s(m_b) \simeq 0.2$
respectively. Next, considerations involving a Coulomb-like potential for the
strong interaction, $V(r) \sim -\alpha_s(1/r)/r$, and the virial theorem lead
to the relation $v \simeq \alpha_s(mv)$. This equation can be solved to
yield $v_c \sim 0.6$ and $v_b \sim 0.3$, pointing therefore to the existence of
a second `small' parameter.

The two small parameters $\alpha_s$ and $v$ suggest a double expansion in
their powers. The theoretical tool to exploit this situation in a theoretically
consistent manner was proposed by Bodwin, Braaten and Lepage in 1994 in the form
of an effective theory, Non-Relativistic QCD (NRQCD) \cite{Bodwin:1994jh}. 
This approach establishes the following factorisation theorem:
\begin{equation}
\sigma[H] = \sum_n \sigma_n(\mu)\langle {\cal O}^H_n(\mu)\rangle \; ,
\end{equation}
i.e. the cross section for the production of the quarkonium state $H$ is given
by a sum of products of a short distance cross section, calculable in
perturbative QCD (pQCD), and a long-distance NRQCD matrix element. The first
term contains the prediction for the production of a heavy quark-antiquark pair
in a given spin, angular momentum and colour (both singlet and octet) state $n$. The
second factor is a non-perturbative contribution accounting for its hadronisation into
the heavy quarkonium state: it cannot be calculated, but NRQCD scaling rules,
in terms of powers of $v$, allow for a controlled truncation of
the series. Finally, $\mu$ is a factorisation scale.

This approach (often mistakenly called the `Colour Octet Model') extends the
Colour Singlet Model (CSM) in a systematic way, in that no divergences should be left
in higher order calculations, and the series can  be meaningfully truncated
according to the scaling rules.

The first phenomenological success of the NRQCD approach was the explanation of
the very large $J/\psi$ and $\psi'$ production at the Tevatron in $p\bar p$
collisions \cite{Braaten:1994vv}: adding colour-octet contributions and fitting
their non-perturbative NRQCD matrix elements to data, their values turned out
to be roughly one-hundredth of the corresponding colour singlet ones (fixed
either by decay rates or by potential models), and therefore in agreement with
the scaling rule $\langle {\cal O}_8^{J/\psi}\rangle /\langle {\cal
O}_1^{J/\psi}\rangle \sim v^4/2N_c \simeq 0.02$, $N_c = 3$ being the number
of colours.

The first solid theoretical success can instead probably be considered the full
next-to-leading order calculation \cite{Petrelli:1997ge} of all the $ij
\to$~quarkonium processes, $i$ and $j$ being a light quark or a gluon. It was
known since the late Seventies that some of these processes develop an
uncanceled singularity in the CSM. In the NRQCD approach this singularity is
seen to be correctly canceled by higher order corrections to the corresponding
colour octet matrix element. One interesting recent development along these
lines is the discovery \cite{Nayak:2005rt} that the definition of the NRQCD
matrix elements must be updated for this property to continue to hold to
next-to-next-to-leading order.

Beyond these first successes, the NRQCD approach has had mixed results. One of
the first problems to surface was the contribution of colour octet channels to
$J/\psi$ photoproduction: with the normalisation fixed by the Tevatron fits,
the theoretical predictions seemed to overshoot \cite{Cacciari:1996dg} the HERA
data  near the elastic region $z=1$. In fact, it was soon argued that in this
region large higher order contributions are likely to be present, and a
meaningful comparison is only possible after a proper all-order resummation.
Once such a resummation was performed, the prediction was indeed found to be in much
better agreement \cite{Beneke:1997qw,Beneke:1999gq,Fleming:2006cd}.

A second instance of a possible discrepancy between NRQCD predictions and
experimental data was found in the exclusive process $e^+e^- \to J/\psi +
\eta_c$. Leading order NRQCD predictions put its cross section around 4--5 fb,
while experimental measurements performed by Belle \cite{Abe:2004ww} and BaBar
\cite{Aubert:2005tj} put it around 20 fb. Also in this case it was however soon
realised that higher order contributions can be important. A recent paper (see
\cite{Bodwin:2006ke} and references therein) puts together a number of such
contributions, and arrives at a prediction of 17.5 $\pm$ 5.7 fb, in good
agreement with the measurements.

The message from these two cases is therefore that when NRQCD predictions are 
to leading order the theoretical uncertainties from missing higher orders can
be huge, partly as a consequence -- especially for charm -- of the relatively
large size of the expansion parameters $\alpha_s(m_c)$ and $v_c$. It is
therefore advisable, when judging how NRQCD fares in comparisons to
experimental measurements,  not to give too much weight to discrepancies with
leading order predictions, and suspend the judgment at least until more
reliable calculations are available.

The third and final kind of comparison between experimental data and NRQCD 
predictions which we address here  revolves around the polarisation of
$J/\psi$'s  produced at the Tevatron. The naive NRQCD result is that, since
most of these particles are predicted to come from gluon fragmentation and the
gluon is nearly real at large transverse momentum, the $J/\psi$ should retain
its transverse polarisation. In fact, more detailed calculations  predict no
net polarisation at low $p_T$, and then a steady increase towards transverse
polarisation as $p_T$ increases. Unfortunately, the data do not seem to support
this picture. Both Tevatron Run I data and preliminary Run II data are rather
compatible with little or no polarisation even at largish $p_T$ (10 -- 20 GeV), 
in contradiction with the prediction (see e.g.  \cite{Lansberg:2006dh} for more
details and references).

There is, at the moment, no clear understanding of what might cause this
discrepancy. If the ``blame'' is to lie on the theoretical side, the NRQCD
velocity scaling rules, which  dictate the relative importance of the higher
orders, might be to blame. Once more, however, one should remember that all the
calculations are leading order ones, and more accurate predictions might once
again change this picture.

\section{Open heavy quark production}

While theoretical predictions for production of closed heavy quark systems like
$J/\psi$ pose evidently an extra level of challenge, it is obvious that they can
be at best only as good as the predictions for open heavy quarks. In this
section we shall therefore explore this issue. The produced system being simpler,
the calculations can be more refined and accurate. On the other hand, the higher
precision leads to the need for extra precautions when studying effects like
non-perturbative fragmentation which, if the overall accuracy were lower, could
be treated much more naively or even almost forgotten altogether.

Modern heavy quark phenomenology tries to provide theoretical descriptions for
observables as close as possible to those which are really measured by
experiments, within realistic phase space acceptances. This means that the
typical process which is calculated looks like the following one:
\begin{equation}
pp\stackrel{pQCD}{\to}Q\stackrel{NP~fragm.}{\to}H_Q\stackrel{decay}{\to}e \; ,
\end{equation}
where $Q$ is the heavy quark, $H_Q$ a heavy hadron, and $e$ represents a generic
final state observable, for instance a lepton. While the decay can be a
complicated process, whose detailed description -- typically via Monte Carlo 
techniques -- is needed to faithfully reproduce the experimental measurement, it
is nevertheless the first part of the reaction, up to the production of the heavy
hadron $H_Q$, which is controlled by QCD and which requires most of our
attention. This process actually contains two different phases, which are not
unambiguously separable: the first is the description in perturbative QCD 
of the production of the heavy quark $Q$ in the hard collision; the second is its
non-perturbative fragmentation into the heavy hadron. They both need to be carefully
studied -- and properly matched -- in order to achieve a good theoretical
precision.

The first building block of heavy quark production calculations is the
factorisation theorem established by Collins, Soper and Sterman 
\cite{Collins:1985gm}, which allows the cross section for production of a heavy
quark to be written, up to higher twist corrections suppressed either by the
heavy quark mass or by the (larger)  transverse momentum, as a convolution of a
calculable short distance perturbative term and of the parton distribution
functions for light partons only. Next-to-leading order (NLO) calculations
based on this theorem were successively performed
\cite{Nason:1988xz,Nason:1989zy,Beenakker:1991ma}, and constitute today the
backbone of all phenomenological predictions.

\begin{figure}[tph]
\begin{center}
\includegraphics[width=6.5cm]{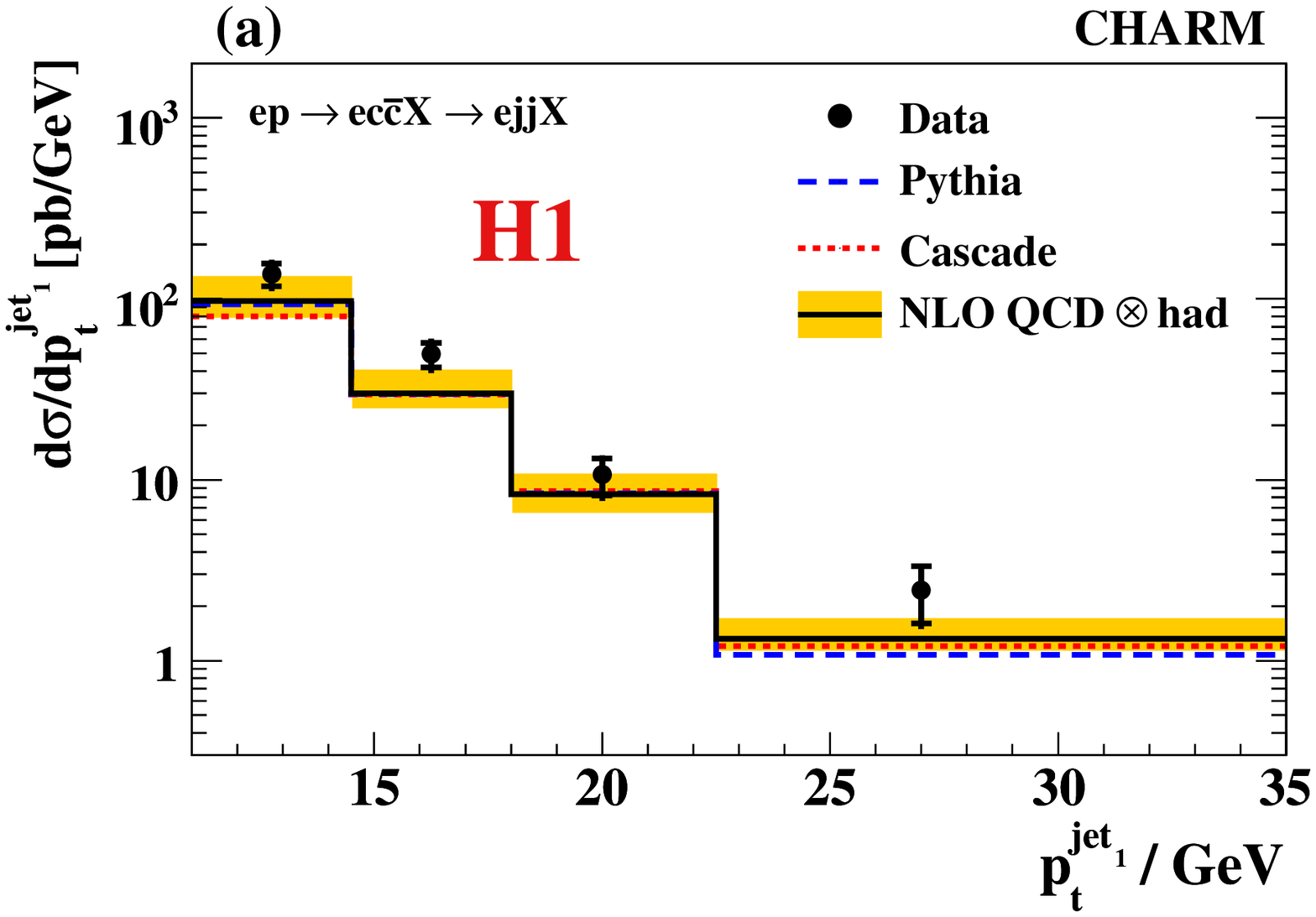}
\includegraphics[width=6.5cm]{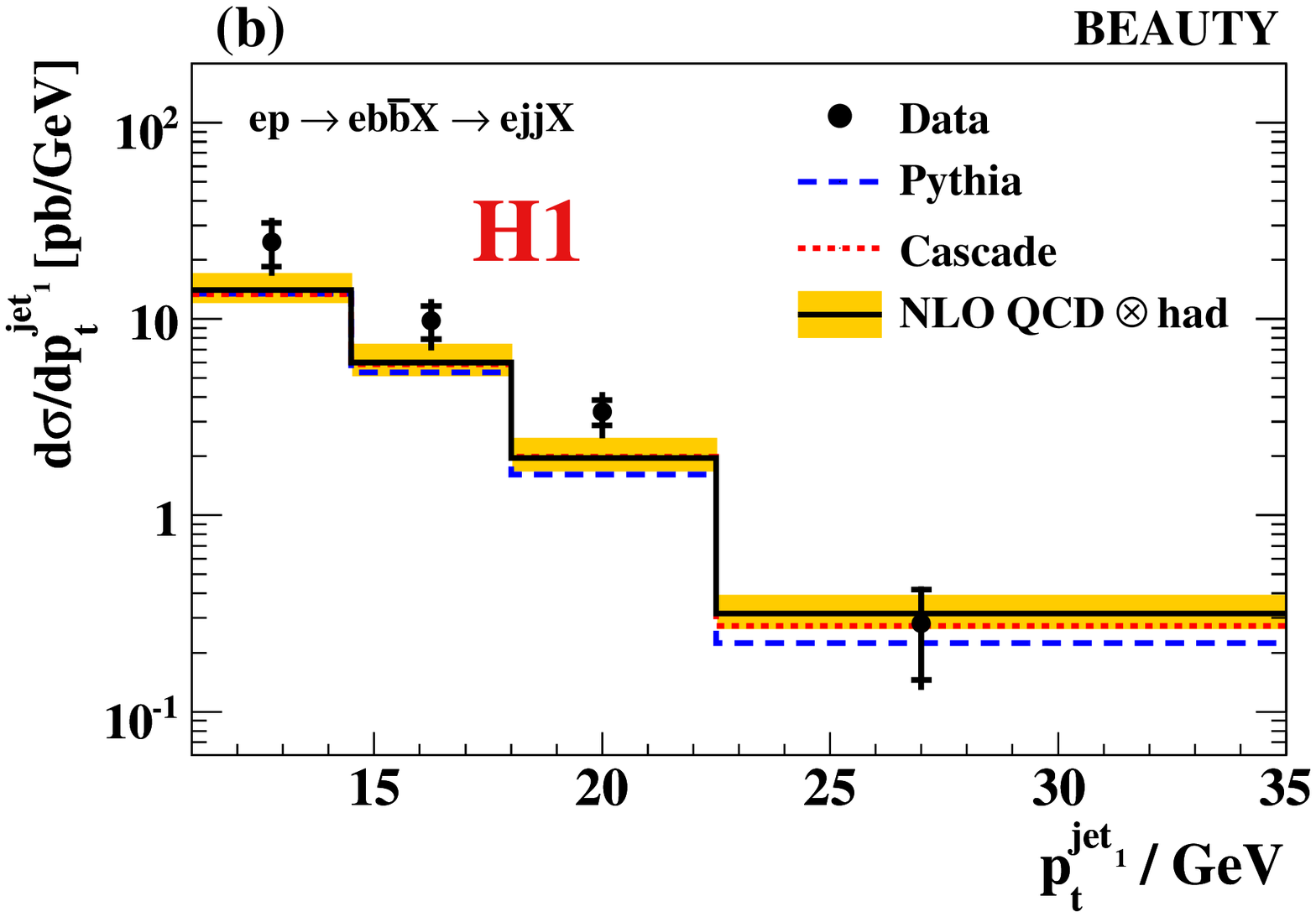}
\includegraphics[width=6.5cm]{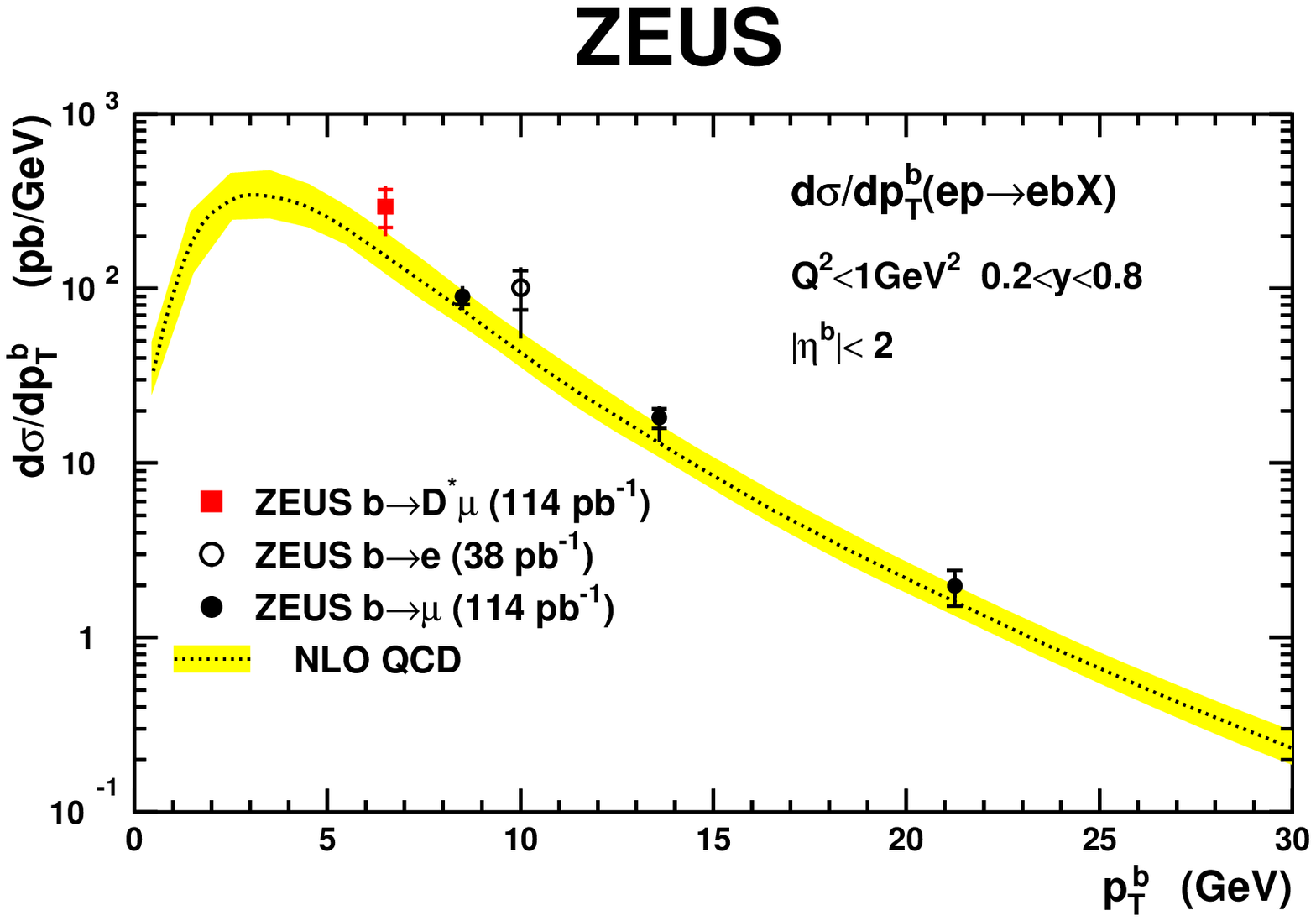}
\includegraphics[width=6cm]{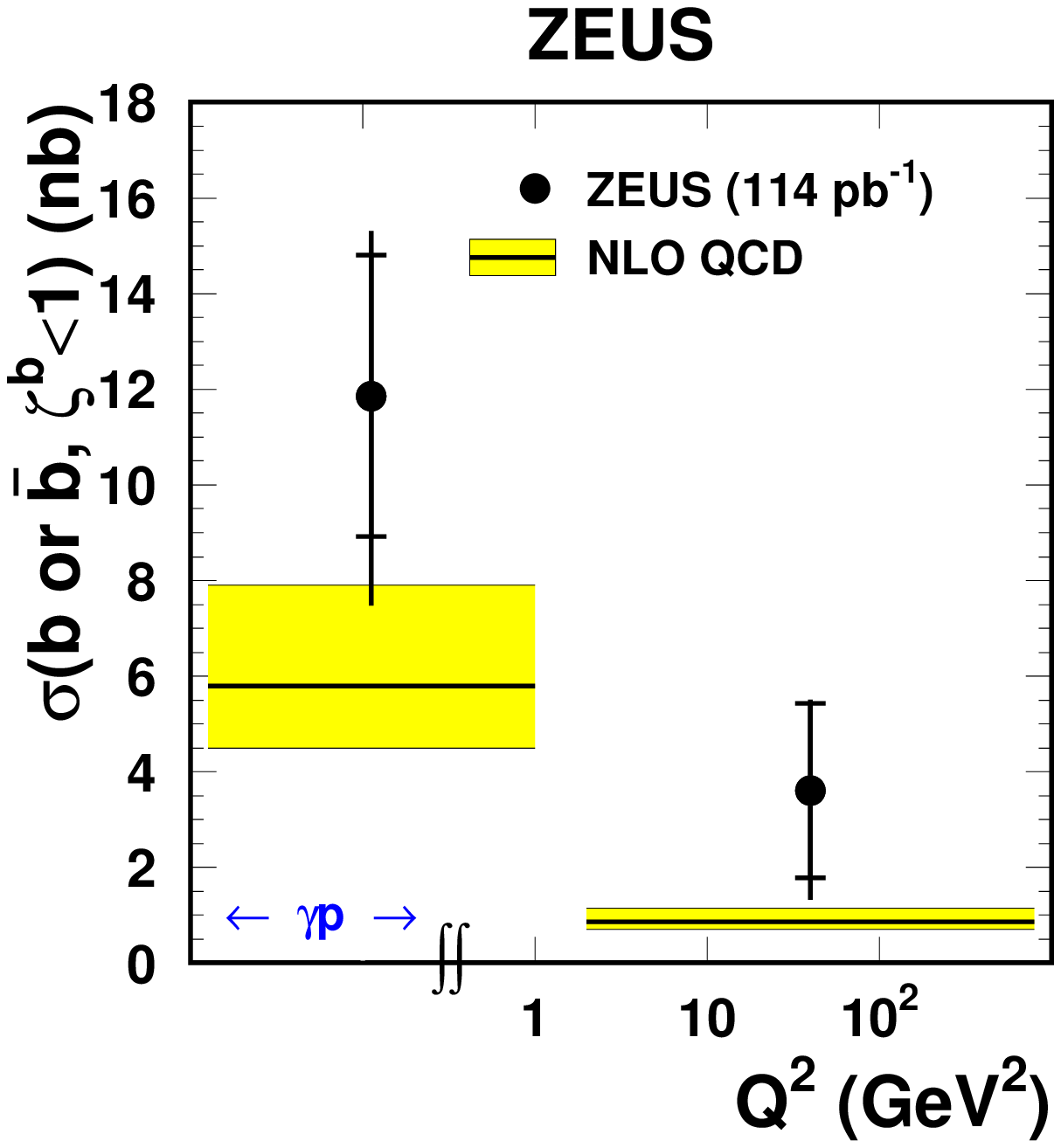}
\includegraphics[width=7cm]{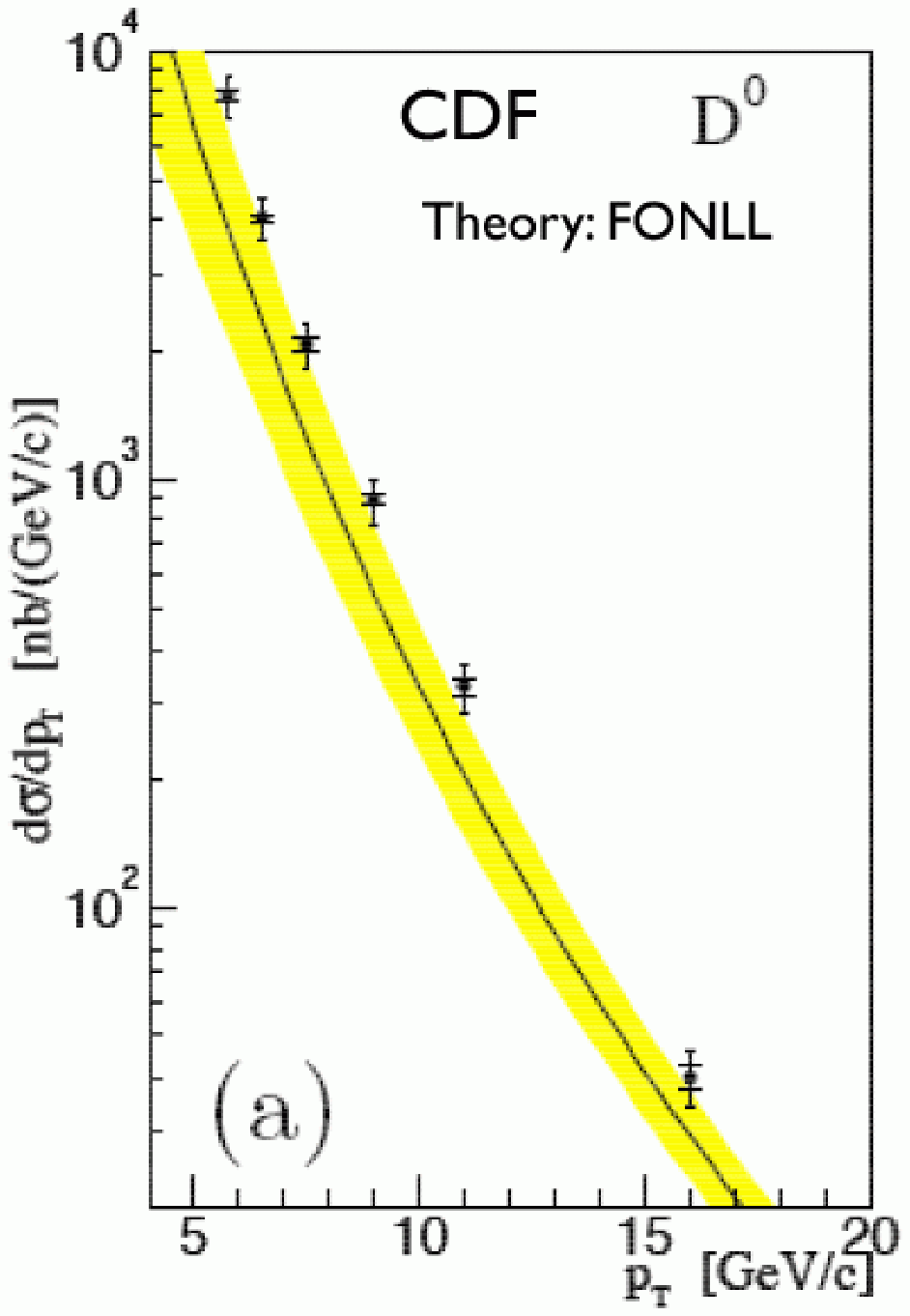}
\includegraphics[width=6.5cm]{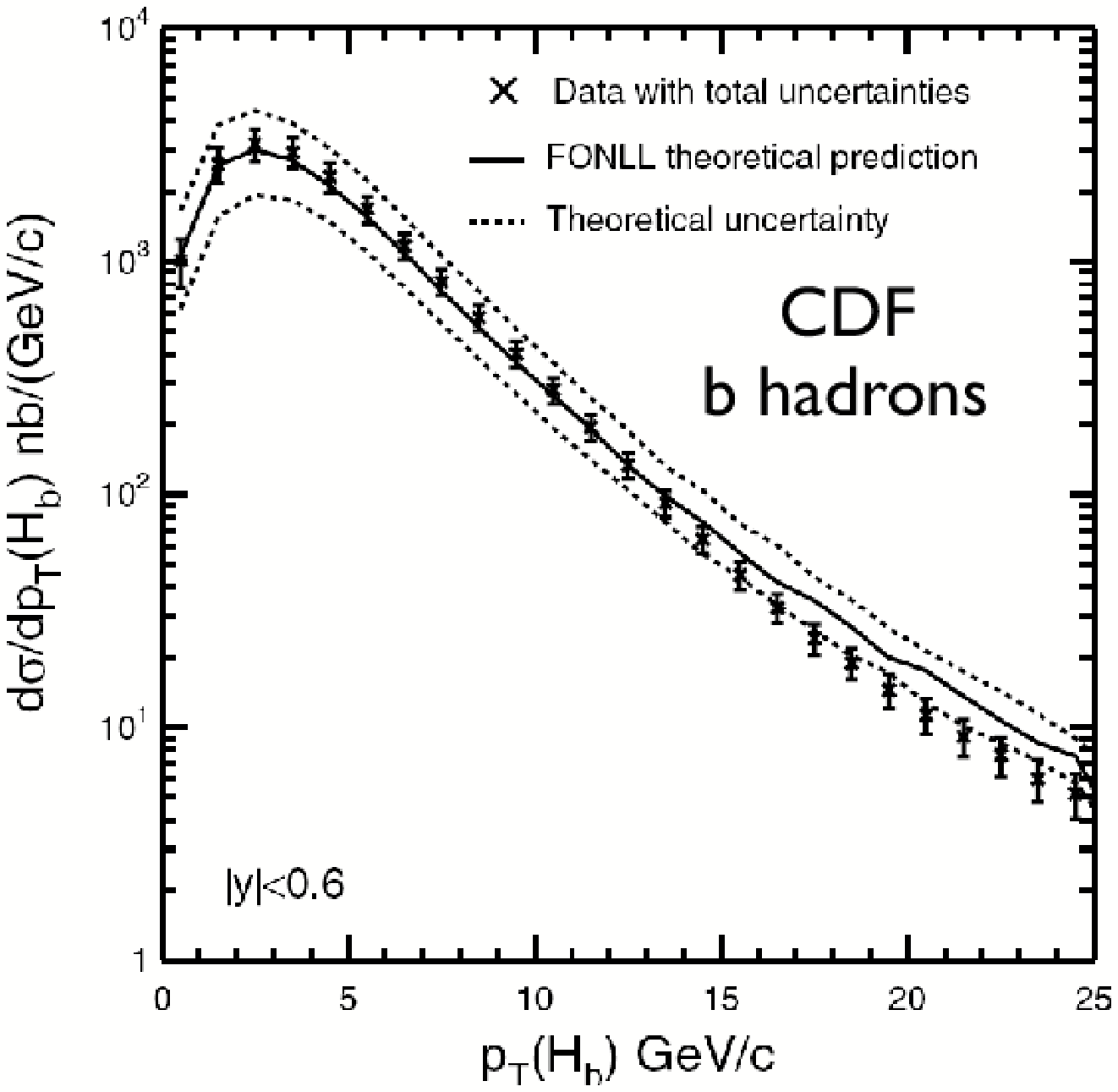}
\end{center}
\caption{\label{manyplots} Some theory vs.data comparisons for a variety of 
exclusive measurements of charm and bottom cross section in different 
experiments. Plots from
\protect\cite{unknown:2006sg,List:2006rs,Acosta:2003ax,Bauer:2006kn}, see
original references therein.}
\end{figure}

While total heavy quark cross sections are a genuine perturbative prediction,
and are usually reliably given by fixed order NLO calculations, for
differential distributions -- closer to what is really measured -- one must
also account for additional effects. On one hand, potentially large 
perturbative logarithms need to be resummed to all orders \cite{Cacciari:1993mq}. On the other, one
needs to consider the momentum degradation related to the hadronisation of the
quarks into the heavy hadrons. This non-perturbative contribution cannot be
calculated, but theoretical arguments show that it is relatively small for
heavy quarks: roughly speaking, momentum degradation is predicted to scale like
${\cal O}(\Lambda/m)$, $\Lambda$ being an hadronic
scale.   Hence, once this contribution is properly defined, extracted from
highly accurate $e^+e^-$ data, and consistently and correctly used in hadronic
collisions, it will not degrade the accuracy provided by the perturbative
calculations, even if its numerical effect will be  non--negligible. One
framework which implements all this is the so--called FONLL \cite{FONLL}, and
many of
the comparisons which will be shown in the following are based on it.

An essential component of a reliable and modern phenomenological prediction is
a sensible estimate of its theoretical uncertainty. Such uncertainty cannot of
course be rigorously established in terms of standard deviations. Rather, it is
usually estimated with a mix of good practice and common sense. The
perturbative calculations contain artificial ingredients, like the
factorisation and renormalisation scale, whose effect on the final result when
varied allows one to estimate its uncertainty. Properly used, and coupled with
experience, this tool can usually provide a reasonably good estimate, provided
of course that the calculation itself is at least minimally reliable (we can't
expect an unreliable calculation to provide a reliable estimate of its own
uncertainty...). One important aspect to keep in mind is that the {\sl
theoretical uncertainty bands} which usually result should not be regarded as
gaussian errors: their `central curve' does not necessarily represent the
`best' prediction. In fact higher order calculations, where available, tend to
disprove this identification. It is therefore wiser to regard an uncertainty
band more like an almost flat probability distribution for the `correct'
prediction, with the probability of it falling within the band depending on the
way it has been obtained, but possibly (again from experience) around 80-90\%
for a typical NLO calculation.

Figure \ref{manyplots} shows some recent comparisons between experimental
results and theoretical predictions (in the form of uncertainty bands) for
production of charm and bottom quarks at various colliders (see
\cite{Lourenco:2006vw} for a recent review of fixed target data). One can see a
generally good agreement within both experimental and theoretical errors. While
the data sometimes tend to  lie on the upper edge of
the theoretical band, they are nevertheless compatible, and moreover a
next-to-next-to-leading
calculation is likely to increase yet a little more the theoretical cross
sections, hence improving the agreement.

\begin{figure}[t]
\begin{center}
\includegraphics[width=0.45\textwidth]{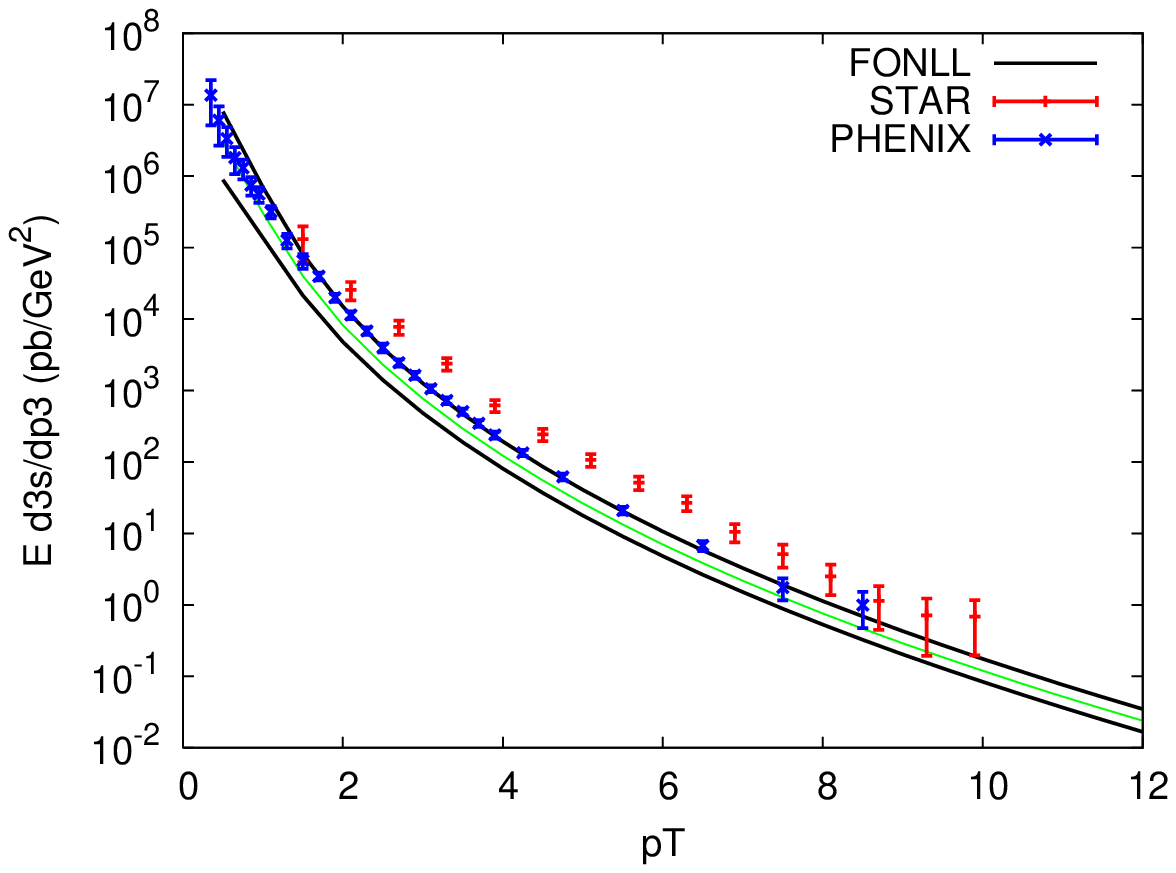}~~~
\includegraphics[width=0.46\textwidth]{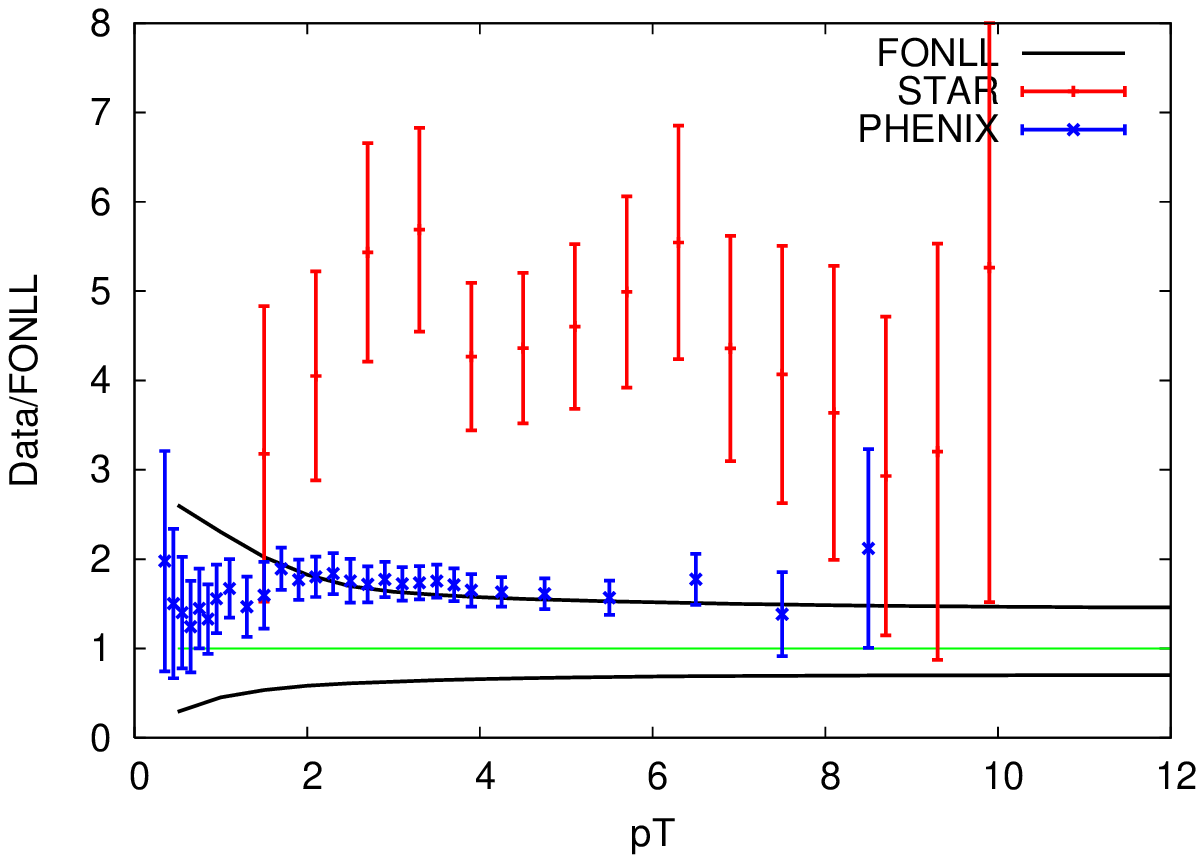}
\caption{\label{rhic} Non-photonic electron cross sections 
from PHENIX and STAR, compared to the FONLL theoretical prediction of
\protect\cite{Cacciari:2005rk}}
\end{center}
\end{figure}

Keeping in mind these comparisons, we consider now the production of charm and
bottom in $pp$ collisions at RHIC, as measured via the transverse momentum
distribution of non-photonic electrons. Both STAR~\cite{Abelev:2006db} and
PHENIX~\cite{Adare:2006hc} have performed this measurement, and FONLL
predictions -- with theoretical simulation of the whole process down to the
decay into electrons -- are available from Ref.~\cite{Cacciari:2005rk}. We can see
from Fig.~\ref{rhic} that while the comparison with PHENIX data looks
qualitatively similar to those previously considered, the STAR cross section 
seems instead to be more significantly larger than the theoretical prediction, 
albeit with large uncertainties.

\section{Conclusions}

NRQCD seems to be a theoretically sound approach to quarkonium production, and
has enjoyed a number of phenomenological successes. Many of the predictions
calculated in this framework are however only leading order ones, and therefore
still subject to large uncertainties. More detailed calculations can certainly
help constrain the theory more. They might also help better understanding
the issue of $J/\psi$ polarisation in $p\bar p$ collisions, presently the
biggest problem for this approach.

As far as calculations for open heavy quark production are concerned, they seem
to do a good job in reproducing most of the experimental measurements for
realistic observables within proper phase space acceptance regions. One might
therefore conclude that even for charm, whose low mass and larger
non-perturbative hadronisation corrections might cast doubts on the predictive
power of the theory, pQCD actually manages to deliver good results. In the
light of a fairly large number of theory--experiment comparisons with
qualitatively similar behaviour, the recent measurement by STAR sticks out as
somewhat  different. More investigations will be necessary  to figure out what
is going on.

\section*{Acknowledgments}

I wish to thank the Organisers of Quark Matter 2006 for the invitation to give
this talk, the financial support, and the superb organisation and the pleasant
atmosphere of this Conference.

\end{document}